\documentclass[letterpaper]{ptephy_v1}
\usepackage{boites}
\usepackage{graphicx}
\usepackage{bm}
\usepackage{amsmath,amssymb,amsfonts}

\newcommand{\nc}{\newcommand}		% new command
\nc{\vc}[1]	{\mbox{\boldmath $#1$}}	% boldmath(vector)
\nc{\bra}       {\langle}               % bra state
\nc{\ket}       {\rangle}               % ket state
\nc{\del}       {\partial}              % bra state
\nc{\AMD}       {{\rm AMD}}
\nc{\TOAMD}     {{\rm TOAMD}}
\newcommand{\lw}[1]{\smash{\lower1.75ex\hbox{#1}}}

\usepackage{color}  % for color  usage: \textcolor{red}{text}
\begin{document}

\title{
High-momentum antisymmetrized molecular dynamics compared with tensor-optimized shell model
for strong tensor correlation}

\author{\name{Takayuki Myo}{1,2}, \name{Hiroshi Toki}{2},  \name{Kiyomi Ikeda}{3}, \name{Hisashi Horiuchi}{2}, \name{Tadahiro Suhara}{4}, \name{Mengjiao Lyu}{2}, \name{Masahiro Isaka}{2}, and \name{Taiichi Yamada}{5}
}
%%%%%%%%%%% The \name command should be used as \name{Insert author name here}{Insert affiliation number here}
%%%%% Please use \thanks for contributed author details

%%%%%%%%%%% The \affil command should be used as \affil{Insert affiliation number here}{Insert author address here}
\address{\affil{1}{General Education, Faculty of Engineering, Osaka Institute of Technology, Osaka, Osaka 535-8585, Japan}\\
\affil{2}{Research Center for Nuclear Physics (RCNP), Osaka University, Ibaraki, Osaka 567-0047, Japan}\\
\affil{3}{RIKEN Nishina Center, Wako, Saitama 351-0198, Japan}\\
\affil{4}{Matsue College of Technology, Matsue 690-8518, Japan}\\
\affil{5}{Laboratory of Physics, Kanto Gakuin University, Yokohama 236-8501, Japan}
\email{takayuki.myo@oit.ac.jp}}

\begin{abstract}%
We treat the tensor correlation in antisymmetrized molecular dynamics (AMD) including large-relative-momentum components among nucleon pairs for finite nuclei.
The tensor correlation is described by using large imaginary centroid vectors of Gaussian wave packets for nucleon pairs with opposite directions, which makes a large relative momentum.
We superpose the AMD basis states, in which one nucleon pair has various relative momenta for all directions; this new method is called ``high-momentum AMD'' (HM-AMD).
We show the results for $^4$He using the effective interaction having a strong tensor force. 
It is found that HM-AMD provides a large tensor matrix element comparable to the case of 
the tensor-optimized shell model (TOSM), in which the two-particle--two-hole (2p-2h) excitations are fully included to describe the tensor correlation.
The results of two methods agree with each other at the level of the Hamiltonian components of $^4$He.
This indicates that in HM-AMD the high-momentum components described by the imaginary centroid vectors of the nucleon pair provide the equivalent effect of the 2p--2h excitations for the tensor correlation.
\end{abstract}

\subjectindex{ D10,  D11}
% D10  Nuclear many-body theories
% D11  Models of nuclear structure

%\parindent0pt

\maketitle
%%%%%%%%%%%%%%%%%%%%%%%%%%%%%
\section{Introduction}
Nuclear force provides a strong short-range repulsion and a strong tensor force in the nucleon--nucleon ($NN$) interaction \cite{pieper01}.
The short-range repulsion reduces the short-range amplitudes of nucleon pairs as short-range correlation.
The tensor force causes strong $S$--$D$ coupling as the tensor correlation involving the $D$-state of nucleon pairs with high-momentum components.
There has been a great deal of work investigating the effects of the correlations induced by the $NN$ interaction in the nuclear wave functions \cite{forest96,suzuki09,feldmeier11,alvioli13}.

Recently, we developed a new variational method for finite nuclei with $NN$ interaction \cite{myo15,myo17a,myo17b,myo17c,myo17d}, in which 
the antisymmetrized molecular dynamics (AMD) is used for nuclear many-body wave functions \cite{kanada03,kanada12}.
We introduce two kinds of variational correlation functions in AMD with tensor-operator and central-operator types.
These correlation functions are successively multiplied to the AMD wave function as the basis states.
We call this variational framework the ``tensor-optimized antisymmetrized molecular dynamics'' (TOAMD).
In TOAMD, we can successively increase the power series of the multiple products of the correlation functions to expand the variational space, although it becomes more time-consuming.

The physical concept of the correlation functions in TOAMD is similar to that of the tensor-optimized shell model (TOSM) \cite{myo05,myo07,myo07_11,myo09,myo11}.
In TOSM, the two-particle--two-hole (2p-2h) excitations are fully included
and play an essential role in describing the strong tensor correlation with large tensor matrix element.
This is due to the spatial shrinkage of the particle states, which brings the high-momentum components in the 2p--2h states \cite{myo05,myo07}.

In the analysis of $s$-shell nuclei with TOAMD \cite{myo17a,myo17b,myo17c}, we obtained results reproducing those of Green's function Monte Carlo using the Argonne $NN$ interaction \cite{kamada01}.  Within the double products of the correlation functions, the numerical accuracy of the binding energy is 80 keV for $^3$H and 1.2 MeV for $^4$He.
This difference can be reduced by increasing the powers of the correlation functions.
In TOAMD, there are two kinds of variational functions; the correlation functions and the AMD wave function.
All the correlation functions are determined independently in the minimization of the total energy, while the AMD wave function is kept as a single configuration. 
In the trial of applying TOAMD to $p$-shell nuclei, we study the AMD wave function with a desire to find an efficient method to reduce the computational time.
One of the promising extensions of TOAMD is to include the multi-configuration of the AMD part, which should efficiently describe the nuclear structure.

So far, various attempts have been made to describe the tensor correlation for nuclei in AMD \cite{dote06,itagaki06,kimura14,itagaki17}.
They tried to extend the AMD wave function to include the contribution of the tensor force in different ways.
However, no results have been obtained to reproduce the large tensor matrix element in the $NN$ interaction, which reaches around $-60$ to $-70$ MeV for $^4$He in the {\it ab initio} calculations \cite{kamada01}.
In AMD, the nucleon wave function has a Gaussian wave packet with a specific centroid position in phase space.
In the description of nuclei, the centroid positions usually have real values and relatively small imaginary ones.
Recently, some groups have introduced the large imaginary values in the centroid positions \cite{kimura14,itagaki17}, corresponding to the high-momentum components of nucleons.
This extension provides an appreciable amount of tensor correlation, but not as large as the {\it ab initio} value.

In this paper, we pursue the idea given by Refs. \cite{kimura14,itagaki17}, and further increase 
the high-momentum components in the AMD wave function as much as possible with the hope of getting the strong tensor correlation. 
To this end, we fully superpose the AMD basis states,
which involve the high-momentum components of two-nucleon pairs with various spin--isospin configurations by putting the imaginary values in possible directions for their Gaussian centroids.
This treatment of the nucleon pair in AMD is shown to correspond to the 2p--2h excitations, which are described in the TOSM framework.
We examine the tensor correlations in the present new AMD approach by comparing them with the results of TOSM.
The TOSM fully treats the 2p--2h excitations and can be used as a criterion for the amount of the tensor correlation.
We show the results of $^4$He in the extended AMD with high-momentum components and in TOSM using the same Hamiltonian and discuss the efficiency of the extended AMD approach.
This study should provide the foundation of the extension of the AMD part in TOAMD for a full account of the nuclear structure of finite nuclei.
%%%%%%%%%%%%%%%%%%%%%%%%%%%%%
\section{Effective interaction}\label{sec:VNN}
We focus on the tensor correlation and use the Hamiltonian with the effective $NN$ interaction $V$ with a tensor force for a mass number $A$ as 

\begin{eqnarray}
    H
&=& T+V~=~\sum_i^A t_i -T_{\rm c.m.}+\sum_{i<j}^A v_{ij},
    \label{eq:Ham}
    \\
    v_{ij}
&=& v_{ij}^{\rm C} + v_{ij}^{\rm T} + v_{ij}^{LS} + v_{ij}^{\rm Coulomb} .
\end{eqnarray}
Here, $t_i$ and $T_{\rm c.m.}$ are the kinetic energy operators of each nucleon and the center-of-mass (c.m.) of $A$-nucleons, respectively.
The central force $v^{\rm C}$ is Volkov No.2 \cite{volkov65} with a Majorana parameter of 0.6.
The $LS$ force $v^{LS}$ is G3RS \cite{tamagaki68,furutani80} with a strength of 900 MeV.  
These forces have often been used in many studies with cluster model and AMD. 
The tensor force $v^{\rm T}$ is the Furutani--Tamagaki force \cite{furutani80,furutani79}, which is based on the bare G3RS potential. 
We have used the same effective interaction in TOSM \cite{myo05,myo07}. 
This interaction overestimates the binding energies of nuclei because the Volkov central force includes the effect of tensor force in the strength. 
In Ref. \cite{itagaki17}, the same interaction is used for the $LS$ and tensor terms.

%%%%%%%%%%%%%%%%%%%%%%%%%%%%%%%%%%%%%%%%%%%%%%%%%%%%%%%%%%%%%%%%%%%
\section{Antisymmetrized molecular dynamics (AMD)}\label{sec:AMD}
We explain the framework of AMD.
The AMD wave function $\Phi_{\rm AMD}$ is a single Slater determinant of $A$-nucleons, given as
\begin{eqnarray}
\Phi_{\rm AMD}
&=& \frac1{\sqrt{A!}}\, {\rm det} \left\{\prod_{i=1}^{A} \phi_i \right\}\,,
\label{eq:AMD}
\\
\phi(\vec r) &=& \left( \frac{2\nu}\pi \right)^{3/4} e^{-\nu(\vec r-\vec Z)^2}\chi_\sigma\chi_\tau.
\label{eq:Gauss}
\end{eqnarray}
The nucleon wave function $\phi( \vec r)$ is a Gaussian wave packet with a range parameter $\nu$
and the centroid position $\vec Z$, which can be a complex number. 
The spin part $\chi_{\sigma}$ is the up ($\uparrow$) or down ($\downarrow$) component for the $z$ direction. 
The isospin part $\chi_{\tau}$ is a proton (p) or neutron (n).
In this study, the range parameter $\nu$ is fixed as 0.25 fm$^{-2}$ and common for all the nucleons. 
We perform the projection of the AMD wave function $\Phi_{\rm AMD}$
on the eigenstates of the total angular momentum $J$ with a quantum number of $M$ and the parity ($\pm$): 
\begin{eqnarray}
\Psi^{J^\pm}_{MK,{\rm AMD}}
&=& P^J_{MK}P^{\pm} \Phi_{\rm AMD}\,,
\label{eq:projection}
\end{eqnarray}
where $P^J_{MK}$ and $P^{\pm}$ are the corresponding projection operators, respectively \cite{kanada12}.

The AMD wave function can be extended to the multi-configuration by applying the generator coordinate method (GCM) using various sets of Gaussian centroids $\{\vec Z_i\}$
with $i=1,\ldots,A$ in the Slater determinant in Eq.~(\ref{eq:AMD}).
When we employ the many basis states of AMD with different sets of $\vec Z$, we superpose these basis states as AMD+GCM.
We express the total wave function $\Psi_{\rm GCM}$ as a linear combination of the AMD basis states as
\begin{eqnarray}
   \Psi_{\rm GCM}
&=& \sum_{\alpha} C_{\alpha}  \Psi_\alpha\,,
   \label{eq:GCM}
\end{eqnarray}
where the label $\alpha$ represents the set of quantum numbers of the projected AMD basis state with a specific set of $\vec Z$.
The generalized eigenvalue problem is solved in Eq.~(\ref{eq:eigen}), and the total energy $E$ and the coefficients $C_\alpha$ are determined.:
\begin{eqnarray}
   \sum_{\beta} \left( H_{\alpha \beta} - E N_{\alpha \beta} \right) C_\beta &=&0,
   \label{eq:eigen}
   \\
   H_{\alpha \beta}
~=~ \langle \Psi_\alpha | H |\Psi_\beta \rangle ,
   \qquad
   N_{\alpha \beta}
&=& \langle \Psi_\alpha | \Psi_\beta \rangle .
\end{eqnarray}

In this study, we focus on the discussion of the tensor correlation of $^4$He in AMD+GCM.
In the usual AMD calculation, $^4$He is described variationally as the $(0s)^4$ configuration with $\vec Z= 0 $ for all nucleons. 
This state corresponds to the 0p--0h state and provides no tensor matrix elements.
We explain how to make the AMD basis states to include the tensor correlation.
According to the previous works \cite{kimura14,itagaki17}, we introduce the imaginary values in the Gaussian centroid position $\vec Z$ of the wave packets in Eq.~(\ref{eq:Gauss}). 
The imaginary value of $\vec Z$ contributes to the single-nucleon momentum as
\begin{eqnarray}
   \frac{\bra \phi | \vec p | \phi \ket}{ \bra \phi | \phi \ket }
   &=& 2\hbar \nu\, {\rm Im}(\vec Z),
\end{eqnarray}
where $\vec p=-i\hbar \nabla$.
This property is utilized to introduce the high-momentum component of the tensor correlation in the AMD wave function.
From the TOSM analysis of light nuclei, it has been shown the importance of the 2p--2h excitations with high-momentum components,
coupled strongly with the 0p--0h states by the tensor force \cite{myo05,myo07,myo07_11}.
We would like to express this 2p-2h effect in AMD and focus on the momenta of two nucleons in nuclei, whose centroid positions are $\vec Z_1$ and $\vec Z_2$ in the wave packets.

In the 0p--0h states, $\vec Z_1=\vec Z_2=0$ for $^4$He with the $(0s)^4$ configuration.
We represent the 2p--2h states by introducing the imaginary values of their positions with opposite signs as 
\begin{eqnarray}
    \vec Z_1&=& i \vec D, \qquad
    \vec Z_2~=~-i \vec D,
    \label{eq:imaginary}
\end{eqnarray}
where the vector $\vec D$ is real and represents the momentum vector of nucleon.
Equation~({\ref{eq:imaginary}) corresponds to the 2p--2h state in the AMD wave function where two nucleons are excited from the $\vec D=0$ state to the non-zero $\vec D$ state.
The vector $\vec D$ is also regarded as the shift of the nucleon wave packet from the origin in the momentum space.
Equation~(\ref{eq:imaginary}) also keeps the c.m. momentum zero for two nucleons, but produces a large relative momentum.
Hereafter we call the two nucleons with this relation a ``{\it high-momentum pair}''.
For $^4$He, we prepare four kinds of high-momentum pairs for spin and isospin as follows:
\begin{eqnarray}
1.~~\mbox{p$_\uparrow$ and n$_\uparrow$}  \,,\qquad
2.~~\mbox{p$_\uparrow$ and n$_\downarrow$}\,,\qquad
3.~~\mbox{p$_\uparrow$ and p$_\downarrow$}\,,\qquad
4.~~\mbox{n$_\uparrow$ and n$_\downarrow$}\,.
\end{eqnarray}
The direction of the high-momentum pair is given by $\vec D$ and we choose two directions; the $z$ direction parallel to that of the intrinsic spins,
and also the $x$ direction as the perpendicular case, which are defined as
\begin{eqnarray}
    z \mbox{ direction} ~:~  \vec D~=~D_z\, \vec e_z,\qquad
    x \mbox{ direction} ~:~  \vec D~=~D_x\, \vec e_x,
\end{eqnarray}
where $\vec e_z$ and $\vec e_x$ are the unit vectors for the $z$ and $x$ directions, respectively.
The lengths $D_z$ and $D_x$ represent the amount of shifts for the corresponding directions. 
In Ref.~\cite{itagaki17}, they consider only one high-momentum proton-neutron pair in the $z$ direction.
Similarly, in the present study, we take one high-momentum pair with all directions among $A$-nucleons in the AMD basis states,
which is suitable to investigate the 2p--2h effect in AMD+GCM.
We take the length of $\vec D$ from 1 fm to 11 fm in steps of 1 fm to include various momenta
in addition to the $\vec D=0$ for the 0p--0h state, and superpose these basis states.
We call this AMD+GCM method ``high-momentum AMD'' (HM-AMD), since the total GCM wave function $\Psi_{\rm GCM}$ with the condition of Eq.~(\ref{eq:imaginary}) contains 
the large relative momentum of the nucleon pair and is able to describe the strong tensor correlation.

We shall discuss the tensor correlation coming from the high-momentum pair in the AMD wave function.
For two nucleons with high-momentum components in Eq.~(\ref{eq:imaginary}),
their relative wave function $\phi_{\rm rel}(\vec r)$ with the relative coordinate $\vec r=\vec r_1-\vec r_2$ is a Gaussian wave packet
with the range parameter $\displaystyle \nu_r=\frac{\nu}{2}$ and the centroid position $\vec Z_{\rm rel}$ as
\begin{eqnarray}
 \phi_{\rm rel}(\vec r) &=& \left(\frac{2\nu_r}{\pi}\right)^{3/4} e^{-\nu_r(\vec r -\vec Z_{\rm rel})^2},
 \qquad
  \vec Z_{\rm rel}
  ~=~ \vec Z_1 - \vec Z_2~=~2i\vec D.
\end{eqnarray}
Then 
\begin{eqnarray}
 \phi_{\rm rel}(\vec r) &=& \left(\frac{2\nu_r}{\pi}\right)^{3/4} 
 e^{-\nu_r \vec r ^2 + 4 i \nu_r \vec D \cdot \vec r + 4 \nu_r \vec D^2}.
	\label{eq:relative}
\end{eqnarray}
The middle term in the exponent of Eq.~(\ref{eq:relative}) produces a plane wave, which can be expanded using the partial wave $\ell$ as
\begin{eqnarray}
 e^{4 i \nu_r \vec D \cdot \vec r}
 = 4\pi \sum_{\ell=0}^\infty  i^\ell j_\ell(4\nu_r D r) \left( Y_\ell(\hat D)\cdot Y_\ell (\hat r) \right).
\end{eqnarray}
This expansion involves the $Y_2(\hat r)$ term, which is able to produce the $D$-state for a high-momentum pair.  With the spin wave functions of the high-momentum pair, this component couples with the $S$-state in the 0p--0h state by the tensor force.
The amount of the $Y_2(\hat r)$ component is adjusted by the shift parameter $D$.
The high-momentum component can also be included in the wave function with a large value of $D$ 
and their suitable combination provides the energetically favored ground state.
In the case of $^4$He ($0^+$), the total orbital angular momentum and the total intrinsic spin both become two at maximum.
In the present HM-AMD, the remaining pair except for the high-momentum pair, is in the $(0s)^2$ state.
This means that partial waves of $\ell=0,~2$ are available for the high-momentum pair.

%%%%%%%%%%%%%%%%%%%%%%%%%%%%%%%%%%%%%%%%%%%%%%%%%%%%%%%%%%%%%%%%
\section{Tensor-optimized shell model (TOSM)}\label{sec:TOSM}
We explain the tensor-optimized shell model (TOSM) for $^4$He \cite{myo05,myo07,myo07_11,myo09,myo11}, which is compared with HM-AMD.
In the concept of TOSM, the tensor force can excite two nucleons in hole states to particle states with  the high-momentum components.
Hence, we include up to the 2p--2h excitations in the configurations of TOSM, 
where we do not truncate the particle states in order to get converging results.

We superpose the 0p--0h, 1p--1h, and 2p--2h configurations in TOSM and define the total wave function $\Psi_{\rm TOSM}$ of TOSM as
\begin{eqnarray}
  \Psi_{\rm TOSM} &=&
  \sum_{k_0} A_{k_0} |{\rm 0p\mbox{-}0h},k_0 \ket
+ \sum_{k_1} A_{k_1} |{\rm 1p\mbox{-}1h},k_1 \ket
+ \sum_{k_2} A_{k_2} |{\rm 2p\mbox{-}2h},k_2 \ket .
\end{eqnarray}
Here, three kinds of the amplitudes, $\{A_{k_0},A_{k_1},A_{k_2}\}$, are determined by the diagonalization of the Hamiltonian matrix in the total-energy minimization. 
The label $k_{i}$ is to distinguish the shell-model configurations.
In the case of $^4$He, the 0p--0h configuration is the $(0s)^4$ one and the 1p--1h and 2p--2h configurations are given as
\begin{eqnarray}
|{\rm 1p\mbox{-}1h},k_1 \ket &=& |(0s)^3({\rm higher}),{k_1} \ket , \qquad
\label{eq:1p1h}
\\
|{\rm 2p\mbox{-}2h},k_2 \ket &=& |(0s)^2({\rm higher})^{2},k_{2}\ket , 
\label{eq:2p2h}
\end{eqnarray}
where ``higher'' means the particle states in higher shells above the $0s$ shell.

We explain the single-particle wave functions in TOSM.
The hole states are expressed by using the harmonic oscillator wave functions,
the length parameter $b$ of which is related to the range parameter $\nu$ in AMD as  $\displaystyle \nu=\frac{1}{2b^2}$.
We set the length $b$ as $\sqrt{2}$ fm, equivalent to the value of $\nu=0.25$ fm$^{-2}$.

We use the Gaussian expansion method for the particle states in the higher shells.
This method has been used in TOSM \cite{myo05,myo07} and various cluster models \cite{horiuchi12,myo14_2}.
We superpose a number of Gaussian basis functions using various length parameters to include the high-momentum components for each particle state. 
In this study, we use nine basis functions at maximum.
The orthonormalized single-particle basis functions are constructed by using a superposition of the Gaussian bases \cite{myo07,myo09}.
The c.m. excitations are eliminated using the Lawson method \cite{lawson}. 
The partial waves of the particle states are taken up to the orbital angular momentum $L_{\rm max}$.
We obtain converging solutions of TOSM with increasing $L_{\rm max}$. 
At this level, the wave function $\Psi_{\rm TOSM}$ fully includes the 2p--2h excitations and the results are compared with the present HM-AMD calculation.

%%%%%%%%%%%%%%%%%%%%%%%%%%%%%%%%%%%%%%%%%%%%
\section{Results}\label{sec:results}

% TOSM/4He/Src10.6
%%%%%%%%%%%%%%%%%%%%%%%%%%%%%%%%%%%%
\begin{figure}[b]
\centering
\includegraphics[width=8.5cm,clip]{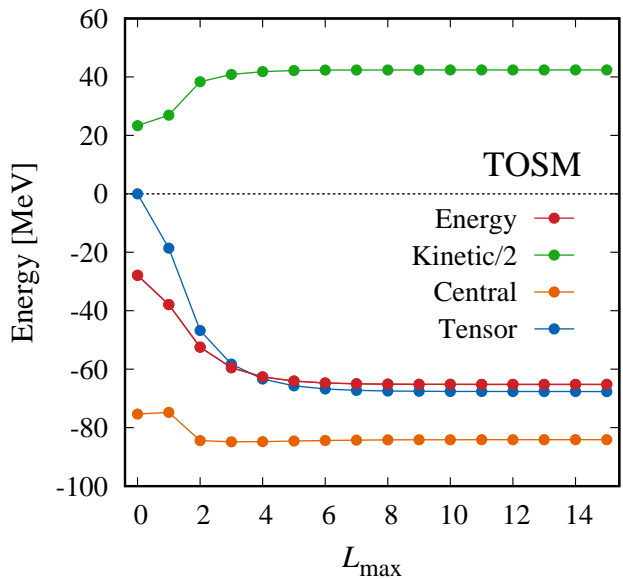}
\caption{Energy convergence of $^4$He in TOSM with respect to the maximum value $L_{\rm max}$ of the partial waves in the configurations. 
The total energy, the kinetic energy with its half value, and the central and tensor matrix elements are shown.}
\label{fig:ene_TOSM}
\end{figure}
%%%%%%%%%%%%%%%%%%%%%%%%%%%%%%%%%%%%

We first explain the TOSM results for $^4$He, since it is important to understand the 2p--2h effect of the tensor force, the amount of the tensor correlation, and the total energy.  
These results should provide the criteria of the tensor correlation produced in HM-AMD.
In Fig.~\ref{fig:ene_TOSM}, we show the convergence of the solutions of TOSM for $^4$He with respect to $L_{\rm max}$. 
We show the total energy and the matrix elements of the Hamiltonian except for the Coulomb and $LS$ terms, which are small.
At around $L_{\rm max}=7$, we clearly obtain the convergence of the solutions. It is found that the convergence for the tensor force is slower than that for the central force.
This indicates the necessity of the higher partial waves of the particle states for the tensor correlation.
Hence, we need a large momentum of $L_{\rm max}/b \sim 5~\mathrm{fm}^{-1}$.
For the particle states, the highest-contributing Gaussian basis functions most commonly have a length parameter of around $0.6\times b$ of the $0s$ hole state \cite{myo07}.
This indicates the spatial shrinkage of particle states inducing the high-momentum components of the tensor correlation in TOSM.
The total energy is $-65$ MeV, which overestimates the experimental value of $-28$ MeV due to the effective interaction.
We obtain the large tensor matrix element as about $-68$ MeV, which represents the 2p--2h effect and is comparable to that of the {\it ab initio} calculation \cite{kamada01}.

% AMD/Src4.5/Data11
%%%%%%%%%%%%%%%%%%%%%%%%%%%%%%%%%%%%
\begin{figure}[b]
\centering
\includegraphics[width=7.3cm,clip]{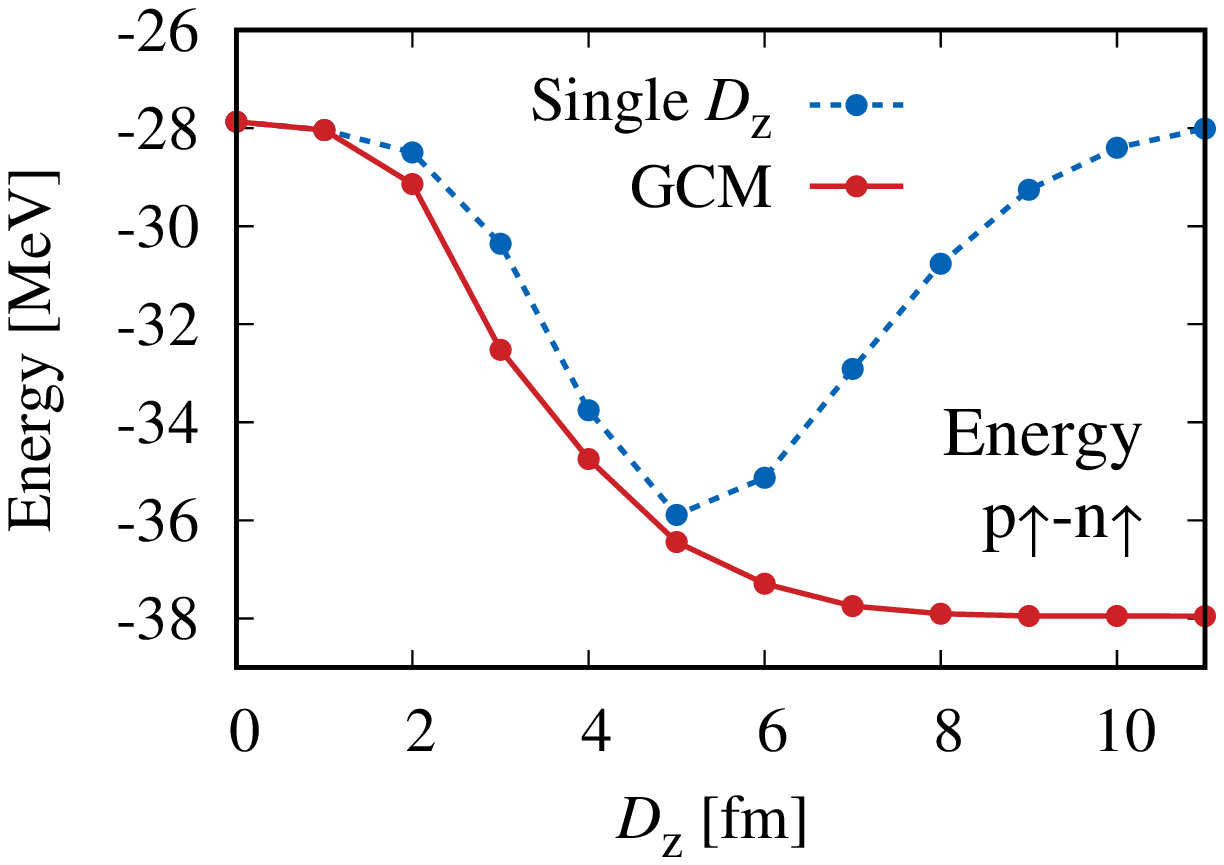}~~~
\includegraphics[width=7.3cm,clip]{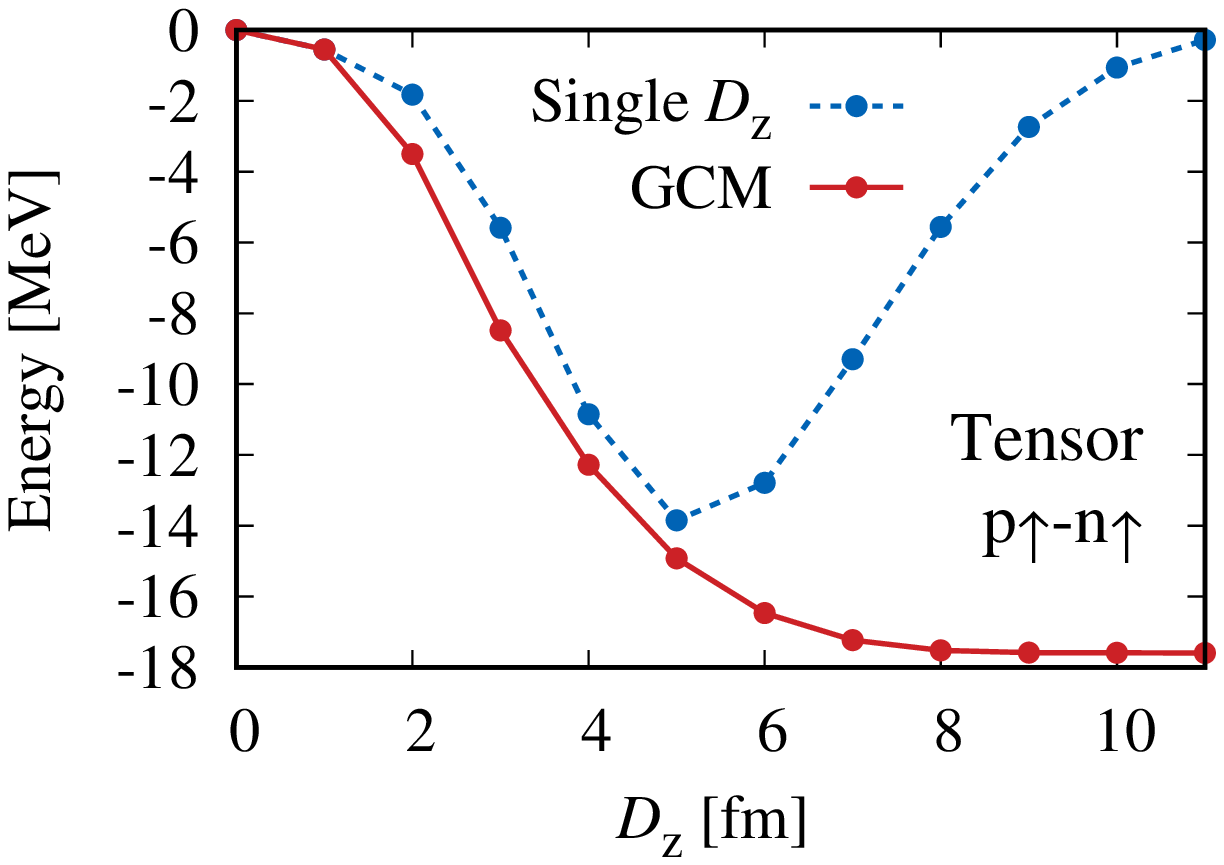}
\caption{Energy (left) and tensor matrix element (right) of $^4$He in AMD+GCM with respect to the imaginary shift parameter $D_z$ for the p$_\uparrow$--n$_\uparrow$ pair. 
Dotted lines are the results using the two basis states with a single $D_z$ and the $(0s)^4$ configuration.
Solid lines are the results of the GCM calculations by successively adding the basis states with increasing $D_z$.}
\label{fig:ene_AMD}
\end{figure}
%%%%%%%%%%%%%%%%%%%%%%%%%%%%%%%%%%%%

We are now in a position to show the results of HM-AMD for $^4$He. 
First we investigate the behavior of solutions as a function of the imaginary-shift vector $\vec D$ in Eq.~(\ref{eq:imaginary}) in the high-momentum pair.
We discuss the case of a p$_\uparrow$--n$_\uparrow$ pair with a shift $D_z$ to the $z$ direction.
In Fig. \ref{fig:ene_AMD}, we plot the energy surface as a function of $D_z$ with dotted lines.
In the calculation, the $(0s)^4$ configuration and the configuration with a specific single $D_z$ for the high-momentum pair are superposed as the two basis states.
This is useful to understand the dependence of the solutions on the shift parameter $D_z$.
On the left-hand side of Fig.~\ref{fig:ene_AMD}, the energy minimum is confirmed clearly at $D_z=5$ fm, which gives the momentum $k_z$ 
of the single nucleon in the high-momentum pair as
\begin{eqnarray}
\bra k_z \ket &=& 2\nu\cdot {\rm Im}(D_z)~=~2.5~{\rm fm}^{-1}.
\end{eqnarray}
This value is large and 1.8 times the empirical Fermi momentum $k_{\rm F}=1.4$ fm$^{-1}$,
indicating that the high-momentum component is variationally favored in the high-momentum pair of AMD.
In Fig. \ref{fig:ene_AMD}, the solid lines show the GCM calculations by successively adding the various $D_z$ components in the AMD basis states as increasing $D_z$ from $D_z=0$ fm. The energy convergence is obtained at around $D_z=10$ fm, corresponding to the momentum of 5 fm$^{-1}$, which agrees with the result of TOSM.

On the right-hand side of Fig. \ref{fig:ene_AMD}, we show the results of the tensor matrix elements under the same calculation conditions as the total energy. 
The overall behavior of the tensor matrix elements is similar to that of the total energy. 
This indicates that the present AMD basis states with high-momentum pairs certainly contribute to a description of the tensor correlation.

%%%%%%%%%%%%%%%%%%%%%%%%%%%%%% 
% AMD :Src4.5/Data09
\begin{table}[t]
\begin{center}
\caption{Energies of $^4$He ($0^+$) in AMD+GCM for four kinds of proton--neutron pairs in units of MeV.
The tensor matrix elements and kinetic energies are also shown.}
\label{tab:AMD_GCM} 
	\begin{tabular}{c|rrrrrrr}
\noalign{\hrule height 0.5pt}
       & \lw{$(0s)^4$} && \multicolumn{2}{c}{$z$}                   && \multicolumn{2}{c}{$x$} \\   \cline{4-5}\cline{7-8}
       &               && $p_\uparrow$-$n_\uparrow$ & $p_\uparrow$-$n_\downarrow$ && $p_\uparrow$-$n_\uparrow$ & $p_\uparrow$-$n_\downarrow$ \\  
\noalign{\hrule height 0.5pt}
Total energy &~$-27.87$ && $-37.95$                & $-38.86$             && $-31.91$                & $-32.02$ \\
Tensor &  $0$     && $-17.59$                & $-21.23$             && $-7.71$                 & $- 6.74$ \\
Kinetic & $46.65$ && $ 57.30$                & $ 61.14$             && $53.68$                 & $ 51.27$ \\
\noalign{\hrule height 0.5pt}
\end{tabular}
\end{center}
\end{table}
%%%%%%%%%%%%%%%%%%%%%%%%%%%%%%

In Table \ref{tab:AMD_GCM}, we show the converged values of four kinds of configurations of high-momentum proton--neutron pairs with for the $z$ and $x$ directions.
It is found that the p$_\uparrow$--n$_\uparrow$ and p$_\uparrow$--n$_\downarrow$ pairs give similar results, providing the large tensor matrix elements for each direction, which is reported for the $z$-direction case in Ref. \cite{itagaki17}.
For the momentum direction, the $z$ direction is favored over the $x$ one.
This can be understood from the property of the tensor force; the tensor force gives the attraction when the relative coordinate of two nucleons is parallel to the intrinsic-spin direction 
due to the tensor operator $S_{12}$ \cite{roth10}.
We also show the kinetic energies, which increase with the inclusion of the high-momentum pair.

% AMD/Src4.5/Data09
%%%%%%%%%%%%%%%%%%%%%%%%%%%%%%%%%%%%
\begin{figure}[b]
\centering
\includegraphics[width=9.0cm,clip]{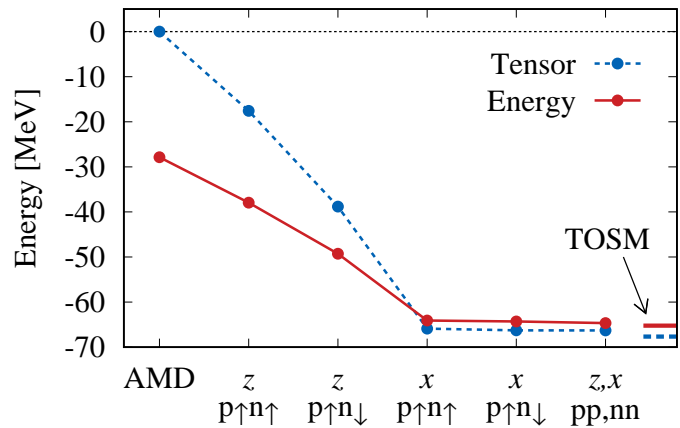}
\caption{Convergence of total energy and the tensor matrix element of $^4$He in HM-AMD by successively adding the high-momentum pairs.
The short horizontal lines on the right-hand side are the results of TOSM.}
\label{fig:AMD_GCM}
\end{figure}
%%%%%%%%%%%%%%%%%%%%%%%%%%%%%%%%%%%%

In Fig. \ref{fig:AMD_GCM}, we finally superpose all the available configurations for high-momentum pairs in AMD+GCM, which is the case of HM-AMD.
It is noted that each AMD basis state involves one high-momentum pair.
We successively add the basis states starting from AMD with the $(0s)^4$ configuration.
In the last term of AMD+GCM as HM-AMD, we add all the basis states of the proton--proton (pp) and neutron--neutron (nn) pairs with the $z$ and $x$ directions, both of which give only the isospin $T=1$ channel.
The final value of the tensor matrix element is $-66$ MeV in HM-AMD, comparable to the {\it ab initio} value \cite{kamada01,myo17a}.
It is found that two kinds of $pn$ pairs in the $z$ direction are necessary to gain the total energy, and one kind of $pn$ pair in the $x$ direction contributes to the solutions. 
For the latter part, the result is almost the same when we first include the p$_\uparrow$-n$_\downarrow$ pair in the $x$ direction.
The contributions of the $pp$ and $nn$ pairs are negligible because of the weaker tensor force in the $T=1$ channel.
For the tensor matrix elements, the value of $-66$ MeV is deeper than the summation of 
the values of each high-momentum pair shown in Table \ref{tab:AMD_GCM} as $-53$ MeV.
This represents the coupling effect by the tensor force among the different configurations of the high-momentum pairs.

We compare the results of HM-AMD with those of TOSM, which are shown on the right-hand side of Fig. \ref{fig:AMD_GCM}.
In Table \ref{tab:AMD_TOSM}, we list the results of the two methods for each Hamiltonian component.
We can confirm a very nice agreement between them. This fact indicates that HM-AMD with one high-momentum pair
provides the high-momentum components and certainly represents the fully 2p--2h excitations induced by the tensor force.
There exist small differences in each Hamiltonian component between the two methods.
One possibility to explain the difference is the 1p--1h excitations that are fully optimized in TOSM, 
while HM-AMD does not include this component in the present AMD basis states.
Another is that the AMD basis states are not the eigenstates of the total isospin $T$. 
The projection to the total $T=0$ state of $^4$He($0^+$) is expected to increase the energy slightly.

%%%%%%%%%%%%%%%%%%%%%%%%%%%%%% 
% AMD :Src4.5/Data09
% TOSM:Src10.6/Data13
\begin{table}[t]
\begin{center}
\caption{Energies of $^4$He ($0^+$) in HM-AMD and TOSM in units of MeV.
The radius is in units of fm.}
\label{tab:AMD_TOSM} 
\begin{tabular}{crr}
\noalign{\hrule height 0.5pt}
        & HM-AMD     &~TOSM~  \\
\noalign{\hrule height 0.5pt}
Total energy  &$-64.66$~~ & $-65.23$~~ \\
Kinetic &$ 83.58$~~ & $ 84.78$~~ \\
Central &$-83.63$~~ & $-84.13$~~ \\
Tensor  &$-66.28$~~ & $-67.66$~~ \\
$LS$    &~~$0.79$~~ & $  0.89$~~ \\
Coulomb &~~$0.88$~~ & $  0.89$~~ \\ \hline
Radius  &~$1.329$~~ & $ 1.320$~~ \\
\noalign{\hrule height 0.5pt}
\end{tabular}
\end{center}
\end{table}
%%%%%%%%%%%%%%%%%%%%%%%%%%%%%%

It is interesting to discuss the spatial structure and the momentum distributions of the tensor correlation \cite{forest96,feldmeier11}.
In TOAMD, the spatial distributions of the correlation functions are shown in Ref. \cite{myo17c}; these are important for understanding the effect of the correlation functions.
In HM-AMD, we will consider a similar analysis to that done in TOAMD including the momentum distributions, which should be shown using a bare $NN$ interaction such as AV8$^\prime$.

%%%%%%%%%%%%%%%%%%%%%%%%%%%%%%%%%%%%%
\section{Summary}\label{sec:summary}
We have developed a new method of ``high-momentum AMD'' (HM-AMD) for the treatment of the tensor correlation in antisymmetrized molecular dynamics (AMD).
We introduce one nucleon pair in the AMD basis state, which has imaginary values for the Gaussian centroid positions in opposite directions and is called a ``{\it high-momentum pair}''.
The high-momentum pair in the AMD wave function can bring the high-momentum components of nucleons in nuclei.
This is the extension of Ref. \cite{itagaki17}, in which they limit the momentum direction parallel to the intrinsic spin.
We additionally include the momentum in the perpendicular direction to the intrinsic spin.
We superpose the AMD basis states with various momentum components for all directions in the GCM calculation as HM-AMD, until we get converging results.

In the results, we obtained a large tensor matrix element, comparable to that of the {\it ab initio} calculation. 
We compare the results of HM-AMD with those of the tensor-optimized shell model (TOSM) using the same Hamiltonian.
In TOSM, the 2p--2h excitations are fully treated in nuclei and the strong tensor correlation is described with high-momentum components. 
This method is suitable for investigating the 2p--2h effect of the tensor force.
The solution of HM-AMD agrees well with that of TOSM for each Hamiltonian component. 
This fact indicates that the AMD basis states with one high-momentum pair can be regarded as 2p--2h excitations involving high-momentum components.

The present analysis focuses on the tensor correlation using the effective central interaction.
It is interesting to apply HM-AMD to nuclei with a bare $NN$ interaction with short-range repulsion as well as the tensor force, 
in which the short-range and tensor correlations should be taken into account simultaneously.
Currently we are investigating the short-range correlation in HM-AMD and we find that the results with one high-momentum pair reproduce those of the tensor-optimized AMD (TOAMD) with a single correlation function,
which will be shown in a forthcoming paper.

In TOAMD, we introduce the correlation functions of tensor-operator and central-operator types, which describe two correlations induced by the $NN$ interaction as mentioned above. In TOAMD, we found that the multi-body terms, which are induced by the intermediate- and long-range tensor force, are extremely important.
TOAMD is considered to be a powerful method to treat the three-body interaction.
The combination of TOAMD and the present high-momentum pairs in HM-AMD is a promising method to describe the multiple correlations in many-body nuclear systems effectively.

%\section*{Acknowledgments}
\ack
This work was supported by the JSPS KAKENHI Grants No. JP15K05091, No. JP15K17662, and No. JP16K05351.

%\vfill\pagebreak
%%%%%%%%%%%%%%%%%%%%%%%%%%%%%%%%%%%%%%%%%%%%%%%%%%%%%%%%%%%%%
%\section*{References}
\nc\PTEP[1]{Prog.\ Theor.\ Exp.\ Phys.,\ \andvol{#1}} %%Added by TD 12/06/14
\nc\PPNP[1]{Prog.\ Part.\ Nucl.\ Phys.,\ \andvol{#1}} %%Added by TD 12/06/14


\begin{thebibliography}{00}
\bibitem{pieper01}    S. C. Pieper and R. B. Wiringa, \JL{Annu. Rev. Nucl. Part. Sci.,51,53,2001}
\bibitem{forest96}    J. L. Forest,  V. R. Pandharipande, S. C. Pieper,d R. B. Wiring, R. Schiavilla, and A. Arriaga, \PRC{54,646,1996}
\bibitem{suzuki09}    Y. Suzuki and W. Horiuchi, \NPA{818,188,2009}
\bibitem{feldmeier11} H. Feldmeier, W. Horiuchi, T. Neff, and Y. Suzuki, \PRC{84,054003,2011}  
\bibitem{alvioli13}   M. Alvioli, C. Ciofi degli Atti, L. P. Kaptari,  C. B. Mezzetti, and H. Morita, \PRC{87,034603,2013} and references therein.
\bibitem{myo15}       T. Myo, H. Toki, K. Ikeda, H. Horiuchi, and T. Suhara, \PTEP{2015,073D02,2015}
\bibitem{myo17a}      T. Myo, H. Toki, K. Ikeda, H. Horiuchi, and T. Suhara, \PLB{769,213,2017} 
\bibitem{myo17b}      T. Myo, H. Toki, K. Ikeda, H. Horiuchi, and T. Suhara, \PRC{95,044314,2017}
\bibitem{myo17c}      T. Myo, H. Toki, K. Ikeda, H. Horiuchi, and T. Suhara, \PTEP{2017,073D01,2017}
\bibitem{myo17d}      T. Myo, H. Toki, K. Ikeda, H. Horiuchi, and T. Suhara, \PRC{96,034309,2017}
\bibitem{kanada03}    Y. Kanada-En'yo, M. Kimura, and H. Horiuchi, \JL{C. R. Phys.,4,497,2003}
\bibitem{kanada12}    Y. Kanada-En'yo, M. Kimura, and A. Ono, \PTEP{2012,01A202,2012}
\bibitem{myo05}       T. Myo, K. Kat\=o, and K. Ikeda, \PTP{113,763,2005}
\bibitem{myo07}       T. Myo, S. Sugimoto, K. Kat\=o, H. Toki, and K. Ikeda, \PTP{117,2007,257}
\bibitem{myo07_11}    T. Myo, K. Kat\=o, H. Toki, and K. Ikeda, \PRC{76,024305,2007}
\bibitem{myo09}       T. Myo, H. Toki and K. Ikeda, \PTP{121,511,2009}
\bibitem{myo11}       T. Myo, A. Umeya, H. Toki, and K. Ikeda, \PRC{84,034315,2011}
\bibitem{kamada01}    H.~Kamada et al.,~\PRC{64,044001,2001} and references therein.
\bibitem{dote06}      A. Dot\'e, Y. Kanada-En'yo, H. Horiuchi, Y. Akaishi, and K. Ikeda, \PTP{115,1069,2006}
\bibitem{itagaki06}   N. Itagaki, H. Masui, M. Ito, S. Aoyama, and K. Ikeda, \PRC{73,034310,2006}
\bibitem{kimura14}    M. Kimura, private communication.
\bibitem{itagaki17}   N. Itagaki and A. Tohsaki, {\tt arXiv:1706.06308[nucl-th]}
\bibitem{volkov65}    A. B. Volkov,~\NP{74,33,1965}
\bibitem{tamagaki68}  R. Tamagaki,~\PTP{39,91,1968} 
\bibitem{furutani80}  H. Furutani et al.,~\PTPS{68,193,1980} 
\bibitem{furutani79}  H.~Furutani, H. Horiuchi, and R. Tamagaki,~\PTP{62,981,1979}
\bibitem{horiuchi12}  H. Horiuchi, K. Ikeda, and K. Kat\=o, \PTPS{192,1,2012}
\bibitem{myo14_2}     T.~Myo, Y. Kikuchi, H. Masui, and K.~Kat\=o, \PPNP{79,1,2014}
\bibitem{lawson}      D. H. Gloeckner, and R.D. Lawson, \PLB{53,313,1974}
\bibitem{roth10}      R. Roth, T. Neff, and H. Feldmeier, \PPNP{65,50,2010}
\end{thebibliography}
\end{document}